\documentstyle[sprocl]{article}

\def\be{\begin{equation}}
\def\ee{\end{equation}}
\def\bea{\begin{eqnarray}}
\def\eea{\end{eqnarray}}
\newcommand{\gapproxeq}{\lower
.7ex\hbox{$\;\stackrel{\textstyle >}{\sim}\;$}}
\newcommand{\lapproxeq}{\lower
.7ex\hbox{$\;\stackrel{\textstyle <}{\sim}\;$}}

\begin{document}

\title{  ARE GLUEBALLS AND HYBRIDS FOUND?}

\author{ F E CLOSE}

\address{Rutherford Appleton Laboratory,\\
Chilton, Didcot, Oxon,
OX11 OQX, England}

\maketitle\abstracts{
The lightest scalar glueball and ground state hybrids may have been found.  I
compare signals reported at this conference with theoretical expectations and
highlight the questions that need to be addressed in forthcoming experiments.}

There has been a sudden and dramatic change of emphasis in the search for
glueballs and hybrids.  In the past a case has been
made on behalf of some signal
or other, believed in strongly
by a handful of theorists or experimentalists and an
uphill battle has been
fought (and usually lost) to convince other than the true
believers that the
holy grail has been found.  At this conference there is a general
belief that glueballs, in particular $f_0$(1500) \cite{cb} and possibly
 $\xi$(2230)~\cite{bepc}
have at last been sighted, and that hybrid states, notably
the
$\pi$(1800) \cite{rep}
and exotic $1^{-+}$(1900) \cite{bnl} also have been found.
I shall attempt to weigh the pros and cons.

\section*{A Scalar Glueball?}

In advance
of these data, theoretical arguments suggested that gluonic activity be
manifested
in the 1.5 GeV  region.  Lattice QCD predicted the lightest ``primitive"
(i.e. quenched approximation) glueball to be $0^{++}$ with mass 1.55 $\pm$ 0.05
GeV \cite{ukqcd}.
Recent lattice computations place the mass slightly higher at
1.74 $\pm$ 0.07 GeV \cite{ibm} with an optimised value for phenomenology
proposed by Teper \cite{teper} of 1.57 $\pm$ 0.09 GeV.  That lattice QCD
computations are now concerned with such fine details represents considerable
progress.
Whatever the final concensus may be, these results suggest that scalar
mesons
in the 1.5 GeV region merit special attention.  If indeed a scalar glueball
exists at such accessible mass, then surely we ought to be able to find it.
Conversely,
if a spectroscopy of glueballs were to be confirmed as predicted by the
lattice, this would have potentially profound implications for the future of
theoretical physics and the numerical simulation of analytically intractable
problems.

Complementing this is the growing evidence that there is now an
overabundance of $0^{++}$ mesons in the $I=0$ channels.  The fact that the
$J^{PC}=0^{++} Q\bar{Q}$ nonet, and the predicted scalar glueball, are in the
same 1.5 GeV
region suggests that there will be mixing between the glueball and the
``conventional"
states; naive expectations about the flavour content of glueball
decays
should therefore be reexamined.  I shall argue that this may be a pivotal
matter in the emerging phenomenology \cite{cafe}.

The $f_0$(1500)
certainly satisfies much of the glueball folklore \cite{fec87}.  Not
only is its mass right but it is seen in production mechanisms that are
traditionally believed to favour glueballs, namely

\noindent 1.  Radiative $J/\psi$ decay \cite{bugs}\\
2.  Central region away from quark beams and targets: $pp\rightarrow p_f(G)ps$
 \cite{aba}\\
3.  $p\bar{p}$ annihilation at low energy where destruction of quarks creates
opportunity for gluons to be manifested \cite{cb}.

At this meeting we have a further tantalising hint in
the sighting \cite{rep} of
$f_0$(1500) in decays of the hybrid meson candidate \cite{cp95,cp951}
 $\pi$(1800)
$\rightarrow \pi f_0$(1500) $\rightarrow \pi\eta\eta$.

The qualitative observation, number 1 above, receives some quantitative
support from ref. \cite{cak,zpli}.  By combining the known B.R.
$(\psi\rightarrow\gamma R$) for any resonance $R$ with perturbative QCD
calculation  of  $\psi\rightarrow\gamma (gg)_R$ where the two gluons are
projected onto the $J^{PC}$ of $R$,
Cakir and Farrar estimate the gluon branching
ratio $B(R\rightarrow gg$).  They suggest that

\begin{equation}
\begin{array}{lcl}
B(R[Q\bar{Q}] \rightarrow gg)& =& 0(\alpha^2_s) \simeq 0.1\nonumber\\
B(R[G] \rightarrow gg)& =& \frac{1}{2} \; \mbox{to} \; 1\nonumber\\
\end{array}
\end{equation}
and illustrate this for known $Q\bar{Q}$ resonances (such as $f_2$(1270)).

The
inferred $B(R\rightarrow gg$) tends to be larger if any of the following occur

\begin{equation}
\begin{array}{ll}
\bullet & B(\psi \rightarrow \gamma R) \;\mbox{is large} \nonumber \\
\bullet & \Gamma (R\rightarrow \mbox{all}) \;\mbox{is small} \nonumber
\\
\bullet & R = 0^{++}  \;\mbox{versus}\; 2^{++} \nonumber \\
\end{array}
\end{equation}

The analysis of ref. \cite{bugs} suggests
$ B(\psi \rightarrow \gamma f_0 (1500)\simeq 10^{-3})$.
 As a rough guide, we find \cite{zpli} if a scalar state around 1500 MeV is
produced at 1ppm,  then $B(S\rightarrow
gg)\simeq$ 90/$\Gamma_T$ (MeV).  Thus a very broad $Q\bar{Q}$ state (width
$\gapproxeq$ 500 MeV) could be present at this level, but for $f_0$(1500) with
$\Gamma_T$ = 100-150 MeV, one infers $B(f_0\rightarrow gg)$ = 0.6 to 0.9
which is far from  $q\bar{q}$.
Such arguments need more
careful study but do add to the interest in the $f_0$(1500).

The width of  $f_0$(1500) is also anomalous for a $^3P_0(Q\bar{Q})$.  For a
$0^{++}$ nonet one expects that
$$
\Gamma (f_0^{n\bar{n}} ) >> \Gamma (a_0) \gapproxeq \Gamma (K^*)
$$
Using data on $2^{++}$ mesons as input one expects the quasi-two body
contributions to be of order 400, 280 and 250 MeV respectively.
 The latter are in
accord with the Crystal Barrel $\Gamma (a_0)\simeq $ 270 $\pm$ 40 and the
$K^*$
width of 287 $\pm$ 23 (essentially all $K\pi$).  The broad $f_0$(1370) could be
the
$n\bar{n}$ state; the $f_0$(1500) width of 116 $\pm$ 16 MeV is far too small
and if $4\pi (\sigma \sigma)$ is a considerable fraction of this, the intrinsic
$\pi\pi, KK, \eta\eta$
will be only tens of MeV.  A detailed discussion is in ref
\cite{cafe}.

Thus
the $f_0$(1500) has the right mass and is produced in the right places to be a
glueball.
Its total width is out of line with expectations for a $Q\bar{Q}$.  Its
branching
ratios are interesting and may also signify a glueball that is mixed in
with the neighbouring $Q\bar{Q}$ nonet.  Whereas gluons decay in a flavour
blind
manner perturbatively, this property will tend to be hidden in strong QCD.
If the flux tube model\cite{ip} is a guide to strong QCD, the decays of
glueballs
will be either into pairs of glueballs (or strongly coupled glue states such
as $\eta$) or by mixing into nearby $Q\bar{Q}$ states of the same quantum
number.
This latter may be expected to be important for $0^{++}$ glueball in the
1.5 GeV region.

We give a pedagogical example that is more general than the particular example
of interest here.

In first order perturbation the mixing of the primitive glueball $G_0$ and
$Q\bar{Q}$ leads to
\begin{equation}
|G) = |G_0) + \xi \left\{ \frac{|u\bar{u}+d\bar{d})}{E(G_0)-E(d\bar{d})} +
\frac{|s\bar{s} )}{E(G_0) - E(s\bar{s})} \right\}
\label{mix}
\end{equation}
where $\xi$ is the mixing amplitude, $E(G_0)$ and $E(Q\bar{Q})$ being the
masses of the relevant states.  In a world where flavour symmetry were exact in
the sense that $E(s\bar{s})\equiv E(d\bar{d})$ then the glueball mixes with $|
u\bar{u} + d\bar{d} + s\bar{s})$ and hence its decays will preserve flavour
symmetry.  Thus we see that in the real world where $E(s\bar{s})\neq
E(d\bar{d})$ dramatic effects may result, especially if the
primitive glueball is in
the midst of the $q \bar{q}$ nonet such that
$$
E(d\bar{d})< E(G_0) < E(s\bar{s})
$$
For example, if the glueball lies midway between these, then the $Q\bar{Q}$
mixture in eqn. 3 is $| u\bar{u} + d\bar{d} - s\bar{s})$ and
the subsequent decays
into meson pairs will violate flavour symmetry radically.
In particular there will be destructive interference in the $K\bar{K}$ channels
arising from the $(u\bar{u} + d\bar{d})$ and the $|s\bar{s})$ components.  In
ref. \cite{cafe} we have discussed this is some detail and suggested that the
 suppression of $K\bar{K}$ observed for $f_0$(1500) is a consequence.

It is
important to note that such a mixing for a pure $Q\bar{Q}$ state  would also
kill $\eta\eta$ along
with $K\bar{K}$.   However, the $|G_0)$ component
can decay into glueball pairs, or into  $\eta\eta$  and $\eta\eta^\prime$ ,
restoring these channels in accord with data.  Indeed, the presence of
$f_0(1500)\rightarrow \eta\eta$  and absence or suppression of $K\bar{K}$ is
itself
supportive of the glueball picture in that for a simple $Q\bar{Q}$ decay, the
$K\bar{K}$ and $\eta\eta$ tend to be highly correlated, independent of the
$Q\bar{Q}$ 1-8    mixing angle.

The experimental agenda  now will be to\\
$\bullet$ establish $f_0$(1450) and $K^*$(1430)\\
$\bullet$ quantify the $KK$ branching ratio of $f_0$(1500)\\
$\bullet$ find the predicted \cite{cafe} $s\bar{s}$ member of the multiplet\\
$\bullet$ clarify status of $f_J$(1720).\\
It is
important to confirm the status of these states in central production and in
$\psi\rightarrow\gamma  X$.  If $f_0$(1500) is the first glueball then the
$2^{++}$ and $0^{-+}$ predicted by the lattice should also be sought.  If the
$\xi$(2230) reported at this  conference  $\psi\rightarrow\gamma X$, is a real
$2^{++}$ state in
 then the beginnings of a glueball spectroscopy
may be at hand.  The production rate may also be quantitatively in accord with
that for a
tensor glueball \cite{zpli}.  Establishing the $2^{++}$ and measuring its
$\eta\eta$ branching ratio is a high priority.

\section*{Quarkballs, Glueloops and Hybrids}

The
origins of the masses of gluonic excitations on the lattice are known only to
the computer.  Those in the flux tube have some heuristic underpinning.  The
$Q\bar{Q}$
are connected by a colour flux with tension 1 GeV/fm which leads to
a linear potential in accord with the conventional  spectroscopy.

The
simplest glue loop is based on four lattice points that are the corners of a
square.
As lattice spacing tends to zero one has a circle, the diameter is $\simeq$
0.5 fm, the circle of flux length is then $\simeq$ 1.5 fm and, at 1 GeV/fm, the
ballpark
1.5 GeV mass emerges.  In the limit of lattice spacing vanishing, its 3-D
realisation is a sphere, and hence it is natural that this is $0^{++}$.

The next simplest configuration is based on an oblong, one link across and two
links long.
The total length of flux is $\simeq \frac{3}{2} $  larger than the square
and the ensuing mass $\simeq  \frac{3}{2} \times $ 1.5 GeV $\simeq $ 2.2 GeV.
In the
3-D continuum
limit this rotates into a rugby ball shape rather than a sphere.  A
decomposition in spherical harmonics contains $L\gapproxeq 0$, in particular
$2^{++}$.
This is by no means rigorous (!) but may help to give a feeling for the
origin of the glueball systematics in this picture, inspired by the lattice.

Finally
one has the prediction that there exist states where the gluonic degrees of
freedom
are excited in the presence of $Q\bar{Q}$.  With the 1 GeV/fm setting the
scale,
one finds \cite{ip} that the lightest of these ``hybrid" states have masses of
order 1 GeV above their conventional $q\bar{q}$ counterparts.  Thus hybrid
charmonium may exist at around 4 GeV, just above the $D\bar{D}$  pair
production
threshold.  More immediately accessible
 are light quark hybrids that are expected
in the 1.5 to 2 GeV range after spin dependent mass splittings are allowed for.

At this conference we have
tantalising sightings of an emerging spectroscopy as I shall now review.

\section*{The Hybrid Candidates}

It is well known that hybrid mesons can have $J^{PC}$ quantum numbers
 in combinations such as $0^{--},0^{+-},
1^{-+}, 2^{+-}$ etc. which are unavailable to conventional mesons and as
such provide a potentially sharp signature.

It was noted in ref.\cite{kokoski85} and confirmed in ref.\cite{cp95}
that the best opportunity for isolating exotic hybrids appears
to be in the $1^{-+}$ wave where, for the I=1 state with mass around 2 GeV,
partial widths are typically

\begin{equation}
\label{bnlwidth}
 \pi b_1 : \pi f_1 : \pi \rho \;
= \; 170 \; MeV : 60 \; MeV : 10 \; MeV
\end{equation}
The narrow $f_1(1285)$ provides a useful tag for the
$1^{-+} \rightarrow \pi f_1$ and ref.\cite{lee94} has recently reported a
signal
in $\pi^- p \rightarrow (\pi f_1) p$ at around 2.0 GeV
that appears to have a resonant phase.

Note
the prediction  that the $\pi \rho$ channel is
not negligible relative to the signal channel
$\pi f_1$
thereby resolving the puzzle of the production
mechanism that was commented upon in ref. \cite{lee94}.
This state may also have been sighted in photoproduction this month\cite{utk}
with $M=1750$ and may be the $X(1775)$ of the Data Tables, ref.\cite{pdg94}.

A recent development
is the realisation that even when hybrid and conventional mesons
have the {\bf same} $J^{PC}$ quantum numbers, they may still be distinguished
\cite{cp95,cp951}
due to their different internal structures which give rise to
characteristic selection rules\cite{pene,ip,cp95}.  As an example consider the
 $\rho(1460)$.

(i) If $q\bar{q}$ in either hybrid or conventional mesons
are in a net spin singlet configuration then the dynamics of the flux-tube
forbids decay into final states consisting only of spin singlet mesons.

For $J^{PC}=1^{--}$  this selection rule
distinguishes between conventional vector mesons
which are $^3S_1$ or $^3D_1$ states and hybrid vector mesons where the
$Q\bar{Q}$ are coupled to a spin singlet. This implies that in the decays of
hybrid $\rho$, the channel
$\pi h_1$ is forbidden whereas $\pi a_1$ is allowed and
that $\pi b_1$ is analogously suppressed for hybrid $\omega$ decays; this
is quite opposite to the case of $^3L_1$ conventional mesons where the
$\pi a_1$ channel is relatively suppressed and $\pi h_1$ or $\pi b_1$
are allowed\cite{busetto,kokoski87}. The extensive analysis of data
in ref.\cite{don2} revealed the clear presence of $\rho(1460)$\cite{pdg94}
 with a strong $\pi a_1$ mode but no sign of $\pi h_1$,
in accord with the hybrid situation. Furthermore, ref.\cite{don2}
finds evidence for $\omega (1440)$ with no visible
decays into $\pi b_1$ which again  contrasts with
 conventional $q\bar{q}$ $(^3S_1$ or $^3D_1$) initial states and in accord
with the hybrid configuration.

(ii) The dynamics of the excited flux-tube in the {\bf hybrid} state
suppresses the decay to mesons where the $q\bar{q}$ are $^3S_1$ or $^1S_0$
$``L=0"$ states. The preferred decay channels are to ($L=0$) + ($L=1$)
pairs\cite{ip,kokoski85}. Thus
in the decays of hybrid $\rho \rightarrow 4\pi$ the $\pi a_1$ content is
predicted to be dominant and the $\rho \rho$ to be absent. The
analysis of ref.\cite{don2} finds such a pattern for $\rho(1460)$.

(iii) The selection rule forbidding ($L=0$) + ($L=0$) final states
no longer operates if the internal structure or size of the two ($L=0$) states
differ\cite{ip,pene}.
 Thus, for example, decays to $\pi + \rho$, $\pi + \omega$ or $K + K^*$
may be significant in some cases\cite{cp95,cp951},
and it is possible that the {\bf production}
strength could be significant where an exchanged
 $\pi, \rho$ or $\omega$ is involved,
as the
exchanged off mass-shell state may have different structure to the incident
on-shell beam particle.
 This may be particularly pronounced in the case
of {\bf photoproduction} where couplings to $\rho \omega$ or $\rho \pi$
 could be considerable when the $\rho$ is effectively replaced
by a photon and the $\omega$ or $\pi$ is exchanged.
This may explain the production of the candidate exotic
$J^{PC}=1^{-+}$ (ref.\cite{lee94}) and a variety of anomalous
signals in photoproduction.

The first calculation of the widths and branching ratios of hybrid
mesons with conventional quantum numbers is in ref.\cite{cp95}:
the
$0^{-+},2^{-+}$ and the
$1^{--}$ are predicted to be potentially accessible.
It is therefore interesting that each of these  $J^{PC}$ combinations
shows rather clear
signals with features characteristic of hybrid dynamics and which do not
fit naturally into a tidy $Q\bar{Q}$ conventional classification.

We have already mentioned the $1^{--}$.
Turning to the $0^{-+}$ wave,  at this conference
that the VES Collaboration at Protvino confirm their enigmatic
and clear $0^{-+}$ signal in diffractive production with 37 GeV
incident pions on beryllium \cite{rep}. Its mass and decays typify those
expected for a hybrid: $M \approx 1790$ MeV, $\Gamma \approx 200$ MeV
in the $(L=0)$ + $(L=1)$ $\bar{q}q$
 channels $\pi^- + f_0; \; K^- + K^*_0, \; K {( K \pi )}_S $ with no
corresponding strong signal in the kinematically allowed $L=0$ two body
channels $\pi + \rho; \; K + K^*$.

The resonance also appears to couple as strongly to
the enigmatic $f_0(980)$ as it does to $f_0(1300)$,
which was commented upon with some surprise in ref. \cite{rep}.
This may be natural for a hybrid at this mass due to the
predicted dominant $KK_0^*$ channel which will feed
the $(KK\pi)_S$ (as observed \cite{rep}) and hence the channel
$\pi f_0(980)$ through the strong affinity of $K\bar{K} \rightarrow f_0(980)$.
Thus the overall expectations for hybrid $0^{-+}$ are in line with
the data of ref.\cite{rep}. Important tests are now that there should be
a measureable decay to the $\pi \rho$ channel with only a small
$\pi f_2$ or $KK^*$ branching ratio.
At this conference we learn that in the $\pi \eta\eta$ final state the glueball
candidate is seen: $\pi(1.8) \rightarrow \pi f_0(1500)
\rightarrow \pi \eta\eta$.
Seeing a glueball in the decays of an excited glue hybrid is suggestive
though it would be nice to see a
Dalitz plot to be sure that this is indeed scalar resonance production and not
a kinematic reflection in the $\pi \eta\eta$ system.

This leaves us with the $2^{-+}$.
There are clear signals of unexplained activity in the
$2^{-+}$ wave in several experiments for which a hybrid interpretation
may offer advantages. These are discussed in ref.\cite{cp95}.

These various signals in the desired channels provide a potentially consistent
picture. The challenge now is to test it. Dedicated high statistics
experiments with the power of modern detection and analysis should re-
examine
these channels.  Ref.\cite{cp951} suggests that the hybrid couplings are
especially
favourable in
{\it low-energy} photoproduction and as such offer a rich opportunity
for the programme at an upgraded CEBAF
 or possibly even at HERA. If the results of ref.\cite{atkinson}
are a guide, then photoproduction may be an important gateway at a range of
energies and the channel $\gamma + N \rightarrow (b_1 \pi) + N$ can
discriminate
hybrid $1^{--}$ and $2^{-+}$ from their conventional counterparts.

Thus to summarise, we suggest that data are consistent with the existence of
low lying multiplets of hybrid mesons based on the mass spectroscopic
predictions of ref.\cite{ip} and the production and decay dynamics of ref.
\cite{cp95}. Specifically the data include

\begin{eqnarray}
 0^{-+} & (1790 \; MeV; \Gamma = 200 \; MeV) & \rightarrow
\hspace{0.2cm} \pi f_0 ; K\bar{K}\pi \\
\nonumber
 1^{-+} & (\sim 2 \; GeV; \Gamma \sim 300 \;  MeV)  & \rightarrow
\hspace{0.2cm}  \pi f_1 ; \pi b_1 (?) \\
\nonumber
 2^{-+} & (\sim 1.8 \; GeV; \Gamma \sim  200 \; MeV)  & \rightarrow
\hspace{0.2cm}  \pi b_1; \pi f_2 \\
\nonumber
 1^{--} & (1460 \; MeV; \Gamma \sim 300 \; MeV)  & \rightarrow
\hspace{0.2cm}  \pi a_1
\end{eqnarray}

Detailed studies of these and other relevant channels are called for together
with analogous searches for their hybrid charmonium analogues, especially in
photoproduction or $e^+ e^-$ annihilation.

\end{document}